\documentclass[twocolumn,showpacs,aps,prl,superscriptaddress]{revtex4}

\usepackage{graphicx}
\usepackage{dcolumn}
\usepackage{epsfig}
\usepackage{amsmath}

\input pubboard/babarsym
\providecommand{\tbline}{\noalign{\vskip 0.05truecm\hrule\vskip0.05truecm}}
\providecommand{\bma}[1]{\boldmath{$#1$}}
\providecommand{\goto}{\rightarrow}
\providecommand{\calB}{\mbox{${\cal B}$}}
\providecommand{\dt}{\deltat}
\providecommand{\sigdt}{\ensuremath{\sigma_{\deltat}}}

\def\s2b{\stwob}
\newcommand{\etal}{{\em et al.}}

\newcommand{\Fpz}{\ensuremath{R_{+/0}}}
\newcommand{\fpz}{\ensuremath{r_{+/0}}}
\newcommand{\bflav}{\ensuremath{B_{\rm flav}}}

\def\dbline{\noalign{\vskip 0.15truecm\hrule}\noalign{\vskip 2pt}\noalign{\hrule\vskip 0.15truecm}}

\newcommand{\skz}{\ensuremath{S_{\etapr\KS}}}
\newcommand{\ckz}{\ensuremath{C_{\etapr\KS}}}

\providecommand{\etapr}{\mbox{$\eta^\prime$}}
\providecommand{\metapr}{\mbox{$m_{\etapr}$}}

\newcommand{\etapKp}{\ensuremath{B^{+}\rightarrow\etapr K^{+}}}

\newcommand{\BetapKp}{\ensuremath{\calB(\etapKp)}}
\newcommand{\retapKp}{\ensuremath{76.9 \pm 3.5 \pm 4.4}}
\newcommand{\RetapKp}{\ensuremath{(\retapKp)\times 10^{-6}}}
\newcommand{\rAetapKp}{\ensuremath{0.037 \pm 0.045\pm 0.011}}
\newcommand{\RAetapKp}{\ensuremath{[-0.04, 0.11]}}
\newcommand{\etapKz}{\ensuremath{\Bz\rightarrow\etapr\Kz}} 
\newcommand{\etapKzs}{\ensuremath{B^{0}\rightarrow\etapr K^0_S}} 
\newcommand{\BetapKz}{\ensuremath{\calB(\etapKz)}}
\newcommand{\retapKz}{\ensuremath{60.6 \pm 5.6 \pm 4.6}}
\newcommand{\RetapKz}{\ensuremath{(\retapKz)\times 10^{-6}}}
\newcommand{\rSetapKz}{\ensuremath{0.02\pm0.34 \pm 0.03}}
\newcommand{\rCetapKz}{\ensuremath{0.10\pm0.22 \pm 0.04}}

\newcommand{\fetapKz}{\ensuremath{\etapr K^0}}
\newcommand{\fetapKzs}{\ensuremath{\etapr K^0_S}}
\newcommand{\fetapKp}{\ensuremath{\etapr K^+}}

\newcommand{\fetapeppKp}{\ensuremath{\etapr_{\eta\pi\pi} K^+}}
\newcommand{\fetaprgKp}{\ensuremath{\etapr_{\rho\gamma} K^+}}
\newcommand{\fetapeppKz}{\ensuremath{\etapr_{\eta\pi\pi} K^0}} 
\newcommand{\fetaprgKz}{\ensuremath{\etapr_{\rho\gamma} K^0}}

\newcommand{\etagg}{\ensuremath{\eta\ra\gaga}}

\newcommand{\etaprd}{\ensuremath{\etapr\ra\eta\pi^+\pi^-}}

\newcommand{\DE}{\ensuremath{\Delta E}}
\newcommand{\mb}{\mes}
\newcommand{\xf}{\mbox{${\cal F}$}}

\newcommand{\costhr}{\ensuremath{\cos\theta_{\rm T}}}
\newcommand{\kzs}{\mbox{$K^0_S$}}
\providecommand{\KS}{\ensuremath{K_S^0}}

\providecommand{\UfourS}{\mbox{$\Upsilon(4S)$}}
\providecommand{\pvec}{{\bf p}}
\providecommand{\half}{\ensuremath{\frac{1}{2}}}

\def\beq{\begin{equation}}
\def\eeq{\end{equation}}
\def\bef{\begin{figure}}
\def\edf{\end{figure}}
\def\ben{\begin{enumerate}}
\def\een{\end{enumerate}}
\def\bear{\begin{array}}
\def\enar{\end{array}}
\def\beqa{\begin{eqnarray}}
\def\eeqa{\end{eqnarray}}

\def\to{\mbox{$\rightarrow$}}

\def\DE{\mbox{${\Delta E}$}}

\def\mes{\mbox{${m_{ES}}$}}

\newcommand{\acp}{\ensuremath{\calA_{ch}}}

\newcommand{\ttag}{\ensuremath{t_{\rm tag}}}

\newcommand{\BaBarYear}    {03}
\newcommand{\BaBarNumber}  {006}

\newcommand{\SLACPubNumber} {9698}

\def\figurebox#1#2#3{%
    \def\arg{#3}%
    \ifx\arg\empty
    {\hfill\vbox{\hsize#2\hrule\hbox to #2{\vrule\hfill\vbox to #1{\hsize#2\vfill}\vrule}\hrule}\hfill}%
    \else
    {\hfill\epsfbox{#3}\hfill}%
    \fi}

\begin{document}

\preprint{\babar-PUB-\BaBarYear/\BaBarNumber} 
\preprint{SLAC-PUB-\SLACPubNumber} 

\begin{flushleft}
\babar-PUB-\BaBarYear/\BaBarNumber\\
SLAC-PUB-\SLACPubNumber\\
\end{flushleft}




\title{
 \large \bf\boldmath Measurements of
 $CP$-violating Asymmetries and Branching Fractions in $B$ Meson Decays
 to $\eta^{\prime} K $ 
}

%
\author{B.~Aubert}
\author{R.~Barate}
\author{D.~Boutigny}
\author{J.-M.~Gaillard}
\author{A.~Hicheur}
\author{Y.~Karyotakis}
\author{J.~P.~Lees}
\author{P.~Robbe}
\author{V.~Tisserand}
\author{A.~Zghiche}
\affiliation{Laboratoire de Physique des Particules, F-74941 Annecy-le-Vieux, France }
% ...
\collaboration{The \babar\ Collaboration}
\noaffiliation

\date{\today}

\begin{abstract}
We present measurements of the branching fractions of the decays
\etapKp\ and \etapKz.  For \etapKzs\ we also measure
the time dependent \CP-violation parameters \skz\ and \ckz, and for
\etapKp\ the time-integrated charge asymmetry \acp. 
The data sample corresponds to 88.9 million \BB\ pairs produced by
\epem\ annihilation at the \UfourS.
The results are $\BetapKp = \RetapKp$, $\BetapKz = 
\RetapKz$, $\skz = \rSetapKz$, $\ckz = \rCetapKz$, and \-$\acp =
\rAetapKp$. 
\end{abstract}


\maketitle

Non-conservation of \CP\ in the neutral $B$ meson system has
been clearly established \cite{babarsin2b,bellesin2b} in decays to
charmonium such as $\Bz\goto J/\psi\KS$.  
The \CP\ effect arises from the interference between mixing and decay
involving the \CP-violating phase $\beta = \arg{(-V_{cd} V^*_{cb}/
V_{td} V^*_{tb})}$ of the Cabibbo-Kobayashi-Maskawa (CKM) mixing matrix,
and appears experimentally as an
asymmetry in the time evolution of the $\Bz\Bzb$ meson pair.
These decays occur via a CKM-favored (though color-suppressed) $b\goto
c$ tree amplitude.

Here we report results of a similar analysis of the decay \etapKzs, a
CKM-suppressed process that is expected to be dominated by penguin
$b\goto s$ transitions, while the tree and electroweak contributions are
expected to be small \cite{sanda,rosner,beneke}.
The observed branching
fraction is 3--10 times larger than initially expected
\cite{sanda}, which has motivated a variety of conjectures by way of
explanation, including flavor singlet
\cite{rosner}\ and charm enhanced \cite{charming}\ terms.
A recent next-to-leading order QCD factorization calculation
\cite{beneke} suggests that the 
decay rate is not significantly enhanced by these mechanisms, but is
adequately predicted by constructive interference between the penguin
diagrams in which the spectator quark is contained in the \etapr\ or in
the kaon.

The results presented in this paper are based on data collected
in 1999--2002 with the \babar\ detector~\cite{BABARNIM}
at the PEP-II asymmetric $e^+e^-$ collider~\cite{pep}
located at the Stanford Linear Accelerator Center.  An integrated
luminosity of 81.9~fb$^{-1}$, corresponding to 
88.9 million \BB\ pairs, was recorded at the $\Upsilon (4S)$
resonance
(center-of-mass energy $\sqrt{s}=10.58\ \gev$).

Charged particles from the \epem\ interactions are detected, and their
momenta measured, by a combination of a vertex tracker (SVT) consisting
of five layers of double-sided silicon microstrip detectors, and a
40-layer central drift chamber, both operating in the 1.5~T magnetic
field of a superconducting solenoid. Photons and electrons are detected
by a CsI(Tl) electromagnetic calorimeter.
Charged particle identification (PID) is provided by the average energy
loss ($dE/dx$) in the tracking devices, and by an internally reflecting
ring imaging Cherenkov detector (DIRC) covering the central region.

From a $\Bz\Bzb$ meson pair produced in \UfourS\ decay we reconstruct one
of the mesons in the final state $f = \fetapKzs$, a \CP\ eigenstate
with eigenvalue $\eta_f=-1$.  For the time evolution measurement, we also
identify the flavor (\Bz\ or \Bzb) and reconstruct the
decay vertex of the partner ($B_{\rm tag}$). 
The asymmetric beam configuration in the laboratory frame
provides a boost of $\beta\gamma = 0.56$ to the $\Upsilon(4S)$, which
allows the determination of the proper decay time difference $\dt \equiv
t_f-\ttag$ from the vertex separation of the two $B$ meson candidates.
The distribution of \dt\ is
\begin{eqnarray}
  F(\dt) &=& 
        \frac{e^{-\left|\deltat\right|/\tau}}{4\tau} [1 \mp\Delta w \pm
                                                   \label{eq:FCPdef}\\
   &&\hspace{-1em}(1-2w)\left( S_f\sin(\deltamd\deltat) -
C_f\cos(\deltamd\deltat)\right)].\nonumber
\end{eqnarray}
The upper (lower) sign denotes a decay accompanied by a \Bz (\Bzb) tag,
$\tau$ is the mean $\Bz$ lifetime, $\deltamd$ is the mixing
frequency, and the mistag parameters $w$ and
$\Delta w$ are the average and difference, respectively, of the probabilities
that a true $\Bz$\,($\Bzb$) meson is tagged as a $\Bzb$\,($\Bz$).
The tagging algorithm is described in 
\cite{babarsin2b}, and has a measured
analyzing power (efficiency times $(1-2w)^2$) of $(28.1\pm0.7)\%$.   

The parameter $C_f$ measures direct \CP\ violation.  If $C_f=0$, then
$S_f=\stwob_{\rm eff}$, with $\beta_{\rm eff}$ equal to $\beta$ combined
with any weak phase difference arising from multiple amplitudes in the
decay.  Assuming the tree amplitudes are negligible, a deviation from
the value found in charmonium channels can be considered an effect of
phases coming from new physics \cite{london}.  Direct \CP\ violation can
also be detected as an asymmetry $\acp =
(\Gamma^--\Gamma^+)/(\Gamma^-+\Gamma^+)$ in the rates
$\Gamma^\pm=\Gamma(B^\pm\ra\etapr K^\pm)$.

We reconstruct a $B$ meson candidate by combining a \Kp\
\cite{conjugates}\
or \KS\ with an \etaprd\ ($\etapr_{\eta\pi\pi}$)
or \etaprrg\ ($\etapr_{\rho\gamma}$).  
The $\KS\ra\pi^+\pi^-$, \etapr, \etagg, and $\rho^0\ra\pi^+\pi^-$
candidates are selected with
requirements on the relevant invariant masses similar to
those of our previous analysis \cite{babarEtapOm}.  Distributions from the
data and from Monte Carlo (MC) simulations \cite{geant}\ guide the choice of
these selection criteria.  For those quantities taken subsequently as
observables for fitting we retain sidebands adequate to
characterize the background as well as the signal.  For charged $B$
decays, the \Kp\ candidate must have an associated DIRC Cherenkov
angle between $-5\,\sigma$ and $+2\,\sigma$ of the value expected for a
kaon.  This requirement rejects 91\% of pions.

The $B$-meson candidate is characterized by the energy substituted mass
$\mes = \sqrt{(\half s + \pvec_0\cdot \pvec_B)^2/E_0^2 - |\pvec_B|^2}$ and
energy difference $\DE = E_B^*-\half\sqrt{s}$, where the subscripts 0 and
$B$ refer to the initial \UfourS\ and the $B$ candidate, respectively,
and the asterisk denotes the \UfourS\ rest frame.  
The resolutions on these
quantities measured for signal events are 29 MeV for \DE\ and $2.9\
\mevcc$ for \mes.
We require $|\DE|\le0.2\ \gev$ and $5.2\le\mes\le5.29\ \gevcc$.

Backgrounds arise primarily from combinatorics among continuum
events. To reject these we make use of the angle
$\theta_T$ between the thrust axis of the $B$ candidate in the \UfourS\
frame and that of the 
rest of the charged tracks and neutral clusters in the event.
The distribution of $\cos{\theta_T}$ is
sharply peaked near $\pm1$ for combinations drawn from jet-like $q\bar
q$ pairs, and nearly uniform for the isotropic $B$ meson decays; we
require $|\cos{\theta_T}|<0.9$.

We obtain the yields and decay time evolution from extended unbinned
maximum likelihood fits, with input
observables \dt, \DE, \mb, $\metapr$, and a Fisher discriminant \xf.
The Fisher discriminant \cite{CLEO-fisher}\ combines four variables: the
angles with respect 
to the beam axis of the $B$ momentum and $B$
thrust axis (in the \UfourS\ frame), and the zeroth and second angular
moments of the energy flow (excluding the $B$ candidate) about the $B$
thrust axis. 

We use MC simulation to estimate backgrounds from other $B$ decays,
including final 
states with and without charm.  These contributions are negligible for
the $\etapr_{\eta\pi\pi}$ modes.  For $\etapr_{\rho\gamma}$ we include
in the fit a \BB\ component (which we find to be small).

Since we measure the correlations among the observables to
be small in the (background-dominated) data samples entering the fit, we
take the probability density function (PDF) for 
each event to be a product of the PDFs for 
the separate observables.  The efficiencies and mistag rates $w$
for each of four tagging categories are measured with a large sample
(\bflav) of 
decays to fully reconstructed flavor eigenstates \cite{babarsin2b}.  
The signatures of the four tagging categories are essentially lepton,
\Kp\ from \Dstar, \Kp, and a flavor-correlated inclusive class.
For each event 
hypothesis $j$ (signal, \BB\ background, continuum background) and
tagging category $k$, we define 
the PDF (to be evaluated with the
observable set for event $i$) as
\begin{equation}
{\cal P}^i_{j,k} =  {\cal P}_j (\mes) \cdot {\cal  P}_j (\DE) \cdot
 { \cal P}_j(\xf) \cdot {\cal P}_j (\metapr) \cdot
 { \cal  P}_j (\dt;\, \sigdt, k)\,.
\end{equation}
The likelihood function for each decay chain is then
\begin{equation}
{\cal L} = \prod_{k} \exp{(-\sum_j Y_{j,k})}
\prod_i^{N_k}\left[\sum_j Y_{j,k} {\cal P}^i_{j,k}\right]\,,
\end{equation}
where
$Y_{j,k}$ is the yield of events of hypothesis $j$ found by the
fitter in category $k$, and $N_k$ is the number of category $k$ events
in the sample.  
  
The signal PDF factor ${ \cal P}_{\rm sig} (\dt;\, \sigdt, k)$ is equal
to the convolution of $F(\dt;\, k)$ (Eq.\ \ref{eq:FCPdef}), with the
signal resolution function, determined from the \bflav\ sample; \sigdt\
is the error on \dt for a given event.
We determine the remaining PDFs from simulation for the
signal and \BB\ background components, and from (\mes,\,\DE) sideband
data for continuum 
background.  Each of the functions ${\cal P}_{\rm sig}(\mes),\ {\cal  P}_{\rm
sig}(\DE),\ { \cal P}_j(\xf),\ 
{\cal  P}_{\rm bkg}(\dt; k)$, and the peaking component of ${\cal
P}_j(\metapr)$ is
parameterized as a Gaussian function, with or without a second or third
Gaussian or 
asymmetric width as required to describe the distribution.  Slowly
varying distributions (combinatoric background under mass or energy
peaks) are represented by linear or quadratic dependences, or
for \mb, by 
the function $x\sqrt{1-x^2}\exp{\left[-\xi(1-x^2)\right]}$,
with $x\equiv2\mes/\sqrt{s}$ and parameter $\xi$.
Large control samples of $B$
decays to charmed final
states of similar topology are used to verify the simulated resolutions
in \DE\ and \mb.

We compute the branching fractions and \acp\ from fits made without \dt\
or flavor tagging.  Seven parameters of the background PDF are free in
the fit, along with signal and continuum background yields, for
$\etapr_{\rho\gamma}K$ the \BB\ background yield, and for charged modes
the signal and background \acp. We compute the branching fractions from
the fitted signal yields, reconstruction efficiencies, daughter
branching fractions, and the number of produced $B$ mesons, assuming
equal production rates of charged and neutral pairs.  To determine the
reconstruction efficiency, including any yield bias of the likelihood
fit, we apply the method to simulated samples constructed to contain the
signal and continuum background populations expected for data.

\begin{figure}[tbp]
 \includegraphics[angle=0,scale=0.4335]{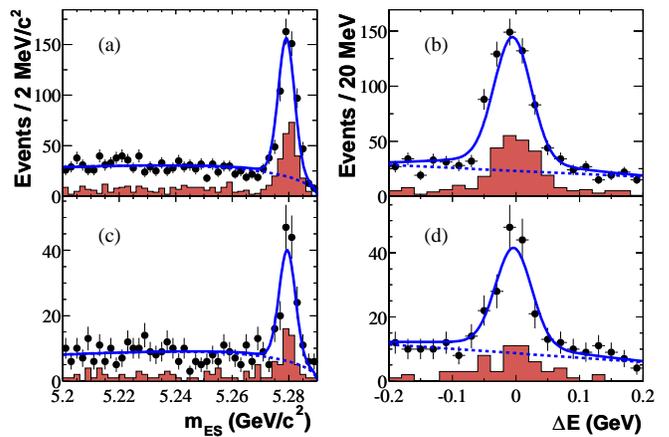}
\vspace{-0.7cm}
 \caption{\label{fig:projMbDE}
 The $B$ candidate \mb\ and \DE\ projections for \etapKp\ (a, b) and
\etapKzs\ (c, d). 
 Points with errors represent the data, solid curves the full fit functions,
 and dashed curves the background functions; the shaded histogram
 represents the  $\eta^\prime_{\eta\pi\pi} K$ subset.  }
\vspace{-.5cm}
\end{figure}

Table \ref{t:results} shows for each decay chain the branching fraction
we measure, together with the quantities entering into its computation.
The purity estimate is given by the ratio of the signal yield to the effective
background plus signal, defined as the square of the error on the yield.
In Fig.\ \ref{fig:projMbDE}\ we show projections onto \mb\ and \DE\ of a
subset of the data for which the signal likelihood
(computed without the variable plotted) exceeds a mode-dependent
threshold that optimizes the sensitivity.

\begin{figure}[tbp]
  \begin{center}
   \includegraphics[scale=0.47]{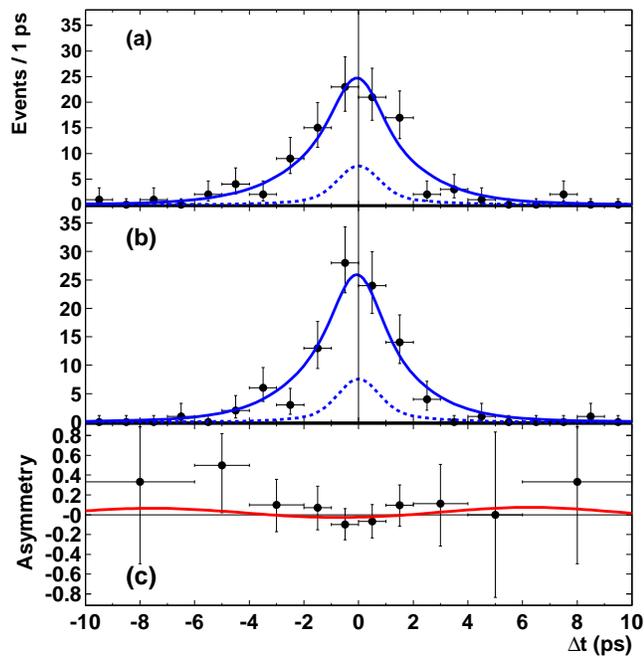}
\end{center}
  \vspace*{-0.5cm}
 \caption{Projections onto $\Delta t$ for \etapKzs\ data (points with errors),
the fit function (solid line), and background function
(dashed line), for (a) \Bz\ and (b) \Bzb\ tagged events, and (c) the
asymmetry between \Bz\ and \Bzb\ tags.}
  \label{fig:DeltaTProj}
\vspace{-.5cm}
\end{figure}

For the time evolution we combine the two decay chains in a single fit
with 28 free parameters: $S_f$, $C_f$, signal fractions (2),
$\etapr_{\rho\gamma} K$ \BB\ background yield (1), common background
\xf\ PDF parameters (3), and separate background \dt,
\mes, \DE, $m_{\etapr}$ PDF parameters (20).  The last four columns of Table
\ref{t:results} give the tagged subsample yields with their 
purity, along with $S_f$ and $C_f$.  The $S_f$ and $C_f$ values for
\etapKp\ are included as a control; they are consistent with zero, as
expected.  We show in Fig.~\ref{fig:DeltaTProj} the $\Delta t$
projections and asymmetry of the combined neutral modes for events
selected as for Fig.\ \ref{fig:projMbDE}.

\providecommand{\bfemsix}{${\cal B} (10^{-6})$}
\providecommand{\msp}{\phantom{-}}
\begin{table*}[btp]
\caption{
Signal yield, purity $P$, detection
efficiency $\epsilon$, daughter branching fraction product that was forced to
100\% in our signal mode simulation, measured branching fraction,
background ($\acp^{qq}$) and signal (\acp) charge asymmetries, tagged
subsample yield $Y_{\rm tag}$ and purity $P_{\rm tag}$, $S_f$,
and $C_f$ for each decay chain, and the combined result for each mode,
with statistical error.}
\label{t:results}
\begin{tabular}{lccccccccccc}
\dbline
Mode		& Yield			& $P$ (\%)	& $\epsilon$ (\%)& $\prod\calB_i$ (\%)	& \bfemsix 	& $\acp^{qq}$ (\%)& \acp (\%)		& $Y_{\rm tag}$	& $P_{\rm tag}$ (\%)	& $S_f$		& $C_f$			\\
\tbline                                                                 					                
\fetapeppKp   & $268\pm19$	& 78			& 25		& 17.4		& $71\pm5$		&$\msp0.6\pm1.6$& $-0.1\pm6.8$		& 183		& 92		& $\msp0.08\pm0.20$	& $-0.16\pm0.15$	\\
\fetaprgKp	& $514\pm31$	& 55			& 24		& 29.5		& $82\pm5$		& $-0.9\pm0.5$  & $\msp6.3\pm5.9$	& 355		& 63		& $-0.07\pm0.16$	& $-0.14\pm0.11$	\\
\bma{\fetapKp}	&			&		&		&		& \bma{76.9\pm3.5}	& $-0.8\pm0.4$  & \bma{3.7\pm4.5}	&		&		& $-0.01\pm0.13$	& $-0.15\pm0.09$	\\
\tbline                                                                 					                
\fetapeppKz	& $48\pm8$		& 75		& 22		& 6.0		& $42\pm7$		&               &			& 31.6		& 75		& $\msp0.75\pm0.51$	& $-0.21\pm0.35$	\\
\fetaprgKz	& $155\pm17$	& 59			& 23		& 10.1		& $76\pm8$		&               &			& 77.6		& 61		& $-0.41\pm0.42$	& $\msp0.24\pm0.27$	\\
\bma{\fetapKz}	&			&		&		&		& \bma{60.6\pm5.6}	&               &			&		&		& \bma{0.02\pm0.34}	& \bma{0.10\pm0.22}	\\

\dbline
\end{tabular}
\vspace{-5mm}
\end{table*}

Most of the systematic errors on yields, which arise from PDF
uncertainties (1--2\%, depending on the decay mode), have already been
incorporated into the overall statistical error, because their
background parameters are free in the fit.  We verify that the
likelihood of each fit is consistent with the distribution found in
simulation. 

The uncertainty in our knowledge of the efficiency is found from
auxiliary studies to be 0.8\%\ per charged track, 2.5\%\ per photon, and
4\%\ per \KS.  Our
estimate of the $B$ production systematic error is 1.1\%.  The estimate
of systematic bias from the fitter itself (0--4\%) comes from fits of
simulated samples with varying background populations.  Published data
\cite{PDG2002}\ provide the $B$ daughter product branching fraction 
uncertainties (3.4\%).
Selection efficiency uncertainties are 1\% for \costhr\ and 0.5\% for PID.
As can be seen in Table \ref{t:results}, the branching fractions we find
for \etapKz\ are rather different (3 standard deviations) as measured
with $\etapr\ra\eta\pi\pi$ or $\etapr\ra\rho\gamma$.  Having exhausted other
explanations, we attribute this difference to a statistical fluctuation,
and include both components in the final measurement.

Using several large inclusive kaon and $B$-decay samples, we find a
systematic uncertainty for \acp\ of 1.1\%\ due to the dependence of
reconstruction efficiency on the charge of the high momentum $K^\pm$.

We find systematic uncertainties for \skz\ and \ckz\ by varying within
their errors the fit parameters controlling the PDF shapes.  We use the
\bflav\ sample to determine the errors associated with the signal \dt\
resolutions, tagging efficiencies, and mistag rates, and published
measurements \cite{PDG2002} for $\tau_B$
and \deltamd.  All of these sum to 0.013 (0.014) for \skz\ (\ckz).  The
contributions from the \mes, \DE, $\metapr$, and \xf\ PDFs are
0.025 (0.014), for \skz\ (\ckz).  We take systematic uncertainties
due to SVT alignment (0.01), beam spot (0.01), boost and $z$ scale
(negligible) from previous determinations of these effects
\cite{babarsin2b}.  We estimate an uncertainty in \ckz\ of 0.025 from
the effect on some tag-side $B$ decays of the
interference between the CKM-suppressed $\bbar\ra \ubar c\dbar$
amplitude with that of the favored $b\ra c\ubar d$ \cite{dcsd}.

We have reconstructed about 800 events of \etapKp\ and 200 of \etapKzs\
with which we have measured the branching fractions, the time-integrated
charge asymmetry \acp\ and the time-dependent asymmetry parameters \skz\
and \ckz.  
We find $\skz = \rSetapKz$ and $\ckz = \rCetapKz$.  These are in
agreement with a previous measurement by the Belle collaboration
\cite{belleEtapkDT}.  A non-zero 
value of \ckz\ would indicate direct \CP\ non-conservation in the
\etapKzs\ decay.  With $\ckz=0$, and provided the decay is
dominated by amplitudes with a single weak phase, \skz\ is equal to
\stwob.  Our result for \skz\
is about two standard deviations smaller than the value obtained with
$B^0\goto J/\psi\kzs$ \cite{babarsin2b,bellesin2b}, and consistent
with zero.

The measured branching fractions are $\BetapKp = \RetapKp$ and $\BetapKz =
\RetapKz$, and we find $\acp = \rAetapKp$.
The null result for \acp\ represents a limit on direct \CP
non-conservation in \etapKp; the 90\% CL limit range is \RAetapKp, and
is consistent with predictions \cite{beneke}.  These values supersede
our previous measurements \cite{babarEtapOm}, and are more than a factor
of two more precise than previous results
\cite{babarEtapOm,cleoBelleEtapkBF}.  The branching fractions depend on
$\Fpz\equiv\BR(\upsbpbm)/\BR(\upsbzbz)$, which we have assumed to be
unity.  To compare the decay rates we form their ratio, making use of
measurements \cite{fpz}\ of $\fpz\equiv\Fpz\times\tau(\Bp)/\tau(\Bz) =
1.14\pm0.06$ (our average); we find $$
\frac{\Gamma(\etapKp)}{\Gamma(\etapKz)} = 1.12 \pm 0.13 \pm 0.06 \pm
0.06, $$
where the last error is from \fpz.

We are grateful for the excellent luminosity and machine conditions
provided by our \pep2\ colleagues, 
and for the substantial dedicated effort from
the computing organizations that support \babar.
The collaborating institutions wish to thank 
SLAC for its support and kind hospitality. 
This work is supported by
DOE
and NSF (USA),
NSERC (Canada),
IHEP (China),
CEA and
CNRS-IN2P3
(France),
BMBF and DFG
(Germany),
INFN (Italy),
FOM (The Netherlands),
NFR (Norway),
MIST (Russia), and
PPARC (United Kingdom). 
Individuals have received support from the 
A.~P.~Sloan Foundation, 
Research Corporation,
and Alexander von Humboldt Foundation.


\begin{thebibliography}{99}
\bibitem{conjugates}
Except as noted explicitly, we use a particle name to denote either
member of a charge conjugate pair.
\end{thebibliography}
\end{document}